\documentclass{aa}

\usepackage{natbib}
\bibpunct{(}{)}{;}{a}{}{,}
\usepackage{amsmath}
\usepackage{graphicx,txfonts}
\usepackage{longtable}
\usepackage{url}
\usepackage{xspace}

\begin{document}
\title{Detection of frequency spacings in the young O-type binary HD\,46149 
from CoRoT photometry\thanks{The CoRoT space
mission was developed and is operated by the French space agency CNES, with
participation of ESA's RSSD and Science Programmes, Austria, Belgium, Brazil,
Germany, and Spain.}$^,$\thanks{Based on observations made with the ESO telescopes
at La Silla Observatory under the ESO Large Programme LP182.D-0356}$^,$\thanks{Based on observations made with the Mercator Telescope, operated on the island of La Palma by the Flemish Community, at the Spanish Observatorio del Roque de los Muchachos of the Instituto de Astrofísica de Canarias.}}

\author{P.~Degroote\inst{1} 
\and M.~Briquet\inst{1}\thanks{Postdoctoral Fellow of the Fund for Scientific
Research, Flanders} 
\and M.~Auvergne\inst{2}
\and S.~Sim\'on-D\'iaz\inst{3,4}
\and C.~Aerts\inst{1,5}
\and A.~Noels\inst{6}
\and M.~Rainer\inst{7}
\and M.~Hareter\inst{8}
\and E.~Poretti\inst{7}
\and L.~Mahy\inst{6}
\and R.~Oreiro\inst{1}
\and M.~Vu{\v c}kovi{\'c}\inst{1}
\and K.~Smolders\inst{1}\thanks{Aspirant Fellow of the Fund for Scientific Research, Flanders.} 
\and A.~Baglin\inst{2}
\and F.~Baudin\inst{9}
\and C.~Catala\inst{2}
\and E.~Michel\inst{2}
\and R.~Samadi\inst{2}
}

\institute{Instituut voor Sterrenkunde, K.U.Leuven, Celestijnenlaan 200D, B-3001
Leuven, Belgium 
\and LESIA, Observatoire de Paris, CNRS UMR 8109, Universit\'e Pierre et Marie Curie, Universit\'e Denis Diderot, 5 place J. Janssen, 92105 Meudon, France
\and Instituto de Astrof\'isica de Canarias, 38200 La Laguna, Tenerife, Spain.
\and Departamento de Astrof\'isica, Universidad de La Laguna, 38205 La Laguna, Tenerife, Spain.
\and Department of Astrophysics, IMAPP, University of Nijmegen, PO Box 9010,
6500 GL Nijmegen, The Netherlands
\and Institut d'Astrophysique et de G\'eophysique Universit\'e de Li\`{e}ge,
All\'{e}e du 6 Ao\^{u}t 17, B-4000 Li\`{e}ge, Belgium
\and  INAF -- Osservatorio Astronomico di Brera, via E. Bianchi 46, 23807 Merate (LC), Italy 
\and Institut f\"ur Astronomie, Universit\"at Wien T\"urkenschanzstrasse 17, A-1180
Vienna, Austria
\and Institut d'Astrophysique Spatiale, CNRS/Universit\'e Paris XI UMR 8617, F-091405 Orsay, France}

\date{Received ? ???? 2010; accepted ? ???? 2010}
\authorrunning{Degroote et al.}
\titlerunning{Pulsations in a O-type star}

\abstract
{}
  {Using the CoRoT space based photometry of the O-type binary HD\,46149, stellar
atmospheric effects related to rotation can be separated from
pulsations, because they leave distinct signatures in the light curve. This offers the
possibility of characterising and exploiting any pulsations seismologically.}
  {Combining high-quality space based photometry, multi-wavelength photometry,
spectroscopy and constraints imposed by binarity and cluster membership, the detected
pulsations in HD\,46149 are analyzed and compared with those for a
grid of stellar evolutionary models in a proof-of-concept approach.}
  {We present evidence of solar-like oscillations in a massive O-type star,
and show that the observed frequency range and spacings are compatible with theoretical
predictions.
Thus, we unlock and confirm the strong potential
of this seismically unexplored region in the HR diagram.}
  {}

\maketitle

%

\section{Introduction}
HD\,46149 is a member of the young open cluster NGC\,2244 in the heart of the
Rosette Nebula, and one of the targets observed by the CoRoT satellite
\citep{baglin2002}, during the short run SRa02 as part of the asteroseismology
programme \citep{michel2006}. There is a general agreement about the
spectral type of HD\,46149: it is catalogued as an O8.5V star
\citep[e.g.][]{jaschek1978,walborn1990}, an O8.5V((f))\footnote{The classification label ((f)) is given when strong He{\sc II} absorption and weak N{\sc III} emission is seen.} \citep[e.g.][]{massey1995}
and as an O8V star \citep[e.g.][]{keenan1985,mahy2009}. The last authors also
confirm that HD\,46149 is actually a binary system with a suspected hot B-type
companion, although their dataset of radial velocity measurements did not suffice
to determine the orbit. Conforming the expectations that a massive main sequence
star of this type does not exhibit a strong stellar wind, \citet{garmany1981}
deduced a weak wind in the primary with a mass loss rate of $\log \dot{M}=-7.7\pm0.3$ from C{\sc IV}
resonance lines in IUE observations.

The known distance of $1.6\pm0.2$ kpc to the cluster and age estimate of 1-6 Myr \citep{bonatto2009},
makes this system a suitable candidate for asteroseismology. Moreover, we can assume that
the chemical composition is similar to that of the cluster,
effectively fixing the metallicity to $Z=0.014$ with solar mixture \citep{asplund2005} and a hydrogen
mass fraction of $X=0.715$ \citep{przybilla2008}. In the past, no serious attempts have
been made to perform asteroseismological modelling of O stars owing to the lack
of detected oscillations. This is primarily
because the pulsation amplitudes are low, and there is possibly contamination by
variable stellar winds. There is some observational evidence of pulsations in
late O-type stars: spectroscopic line profile variations with amplitudes of
$\sim 5$\,km\,s$^{-1}$ have been detected and connected to nonradial pulsations
in, e.g., $\xi$\,Persei and $\lambda$\,Cephei \citep{dejong1999}. Also in photometry, variations have been detected and likely related to pulsation \citep[e.g.,][]{walker2005,rauw2008}. However, possible
cyclical modulation of the wind makes the search for nonradial pulsations particularly
troublesome, since the relevant frequency domains overlap. It is still an open question
whether these mass outflows are connected to pulsations, to a
nonhomogeneous magnetic field at the surface of the star \citep{hubrig2008}, or to 
yet another phenomenon.

The low spectroscopic amplitudes translate to photometric amplitudes below mmag level,
which is difficult to detect from the ground, but is within reach of CoRoT's high-precision space-based photometry. The high duty cycle is fit for monitoring full rotation cycles,
capturing both the influence of the wind and pulsations.

In addition, a spectroscopic campaign has been set up involving 3 different telescopes, partially
overlapping with the CoRoT observations and continuing over the months following
the CoRoT run, to find clues to the open questions in the field of O-type star
pulsations and stellar winds. The observations are summarised in Table\,\ref{tbl:speclog}.

In the following sections, we use the grid of non-rotating stellar models computed by one of us (MB) with the evolutionary code CL\'ES \citep[Code Li\'egeois d’\'Evolution Stellaire, ][]{scuflaire2008}, for interpretating the O9V star HD46202. We used the OPAL2001 equation of state \citep{rogers2002,caughlan1988}, with nuclear reaction rates from \citet{formicola2004} for the $^{14}$N$(p,\gamma)^{15}$ cross-section. Convective transport is treated by using the classical mixing length theory of convection \citep{bohm1958}). For the chemical composition, we used the solar mixture from \citet{asplund2005}. We used OP opacity tables \citet{seaton2005} computed for this mixture. These tables are completed at $\log \rm{T} < 4.1$ with the low-temperature tables of \citet{ferguson2005}. In the calculations, a static atmosphere and no mass loss were assumed.

We fixed the metallicity to $Z = 0.014$ and the hydrogen mass fraction to $X = 0.715$, according to the values derived for the cluster, and in agreement with the solar neighbourhood (Przybilla et al. 2008). We considered a mass range between 20 and 28\,M$_\odot$ in steps of 0.1\,M$_\odot$  and overshoot parameters $\alpha_{\rm ov}$ = 0.0 −- 0.5 pressure scale heights in steps of 0.05. To better match the observed binary system, the grid was extended to 35\,M$_\odot$ in steps of 0.5 M$_\odot$ and an overshoot parameter of $\alpha_{\rm ov}=0.2$.

Afterwards, for each main-sequence stellar model, we calculated the theoretical frequency spectrum of low-order p- and g-modes with a degree of the oscillation up to $\ell = 4$ using a standard adiabatic code for non-rotating stellar models \citep{scuflaire2008b}). In Sect.\,\ref{sec:dataanalysis}, we use the CoRoT light curve to analyse the long- and short-period variability. In Sect.\,\ref{sec:orbital} we use the spectra to characterise the binary system to determine the fundamental parameters of the components, and end with a remark on the absence of spectroscopic variability.

Given all observational constraints, we finally interpret the short- and long-period variations in the CoRoT light curve in Sect.\,\ref{sec:discussion} in terms of rotation and stochastically excited pulsations, respectively.

\section{The CoRoT data}\label{sec:dataanalysis}

\subsection{Frequency analysis}

\begin{figure}
\includegraphics[width=\columnwidth]{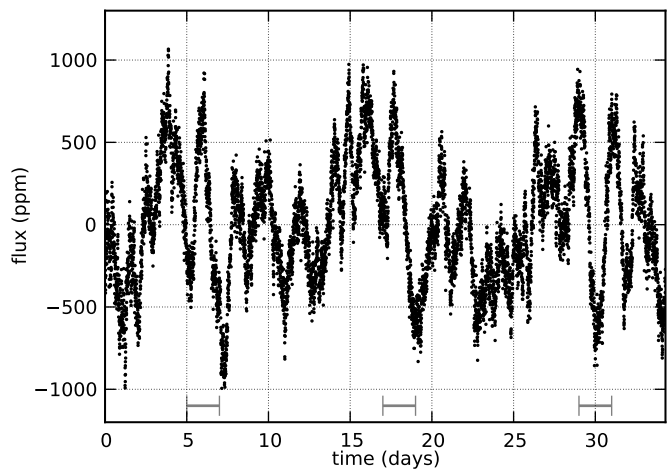}
\caption{Light curve of HD\,46149 from CoRoT space based photometry, binned per 10
observations. The times of
occurrence of the features causing the highest amplitude variation are indicated in grey ($t_0 = 2454748.48856$\,HJD).}
 \label{fig:lc}
\end{figure}

The CoRoT satellite observed HD\,46149 from 2454748.48856 HJD for 34 days with a sampling rate of 32\,s (Fig.\,\ref{fig:lc}). In the frequency spectrum of the reduced light curve\footnote{N2 level data available at \url{http://idoc-corot.ias.u-psud.fr}}, we still see some instrumental effects, and distinguish the large-amplitude, low-frequency signal from the low-amplitude high-frequency signal. In the following, we first remove most of the instrumental signal. Then, the low-frequency signal is separated from the high-frequency signal to analyse them separately, mainly focusing on the latter.

All flagged observations in the light curve were removed. The decreasing
trend in the CoRoT light curve was removed by dividing by a linear fit, since it is
visible in almost all CoRoT targets and is thus considered to be of instrumental
origin \citep{auvergne2009}.

\begin{figure}
\includegraphics[width=\columnwidth]{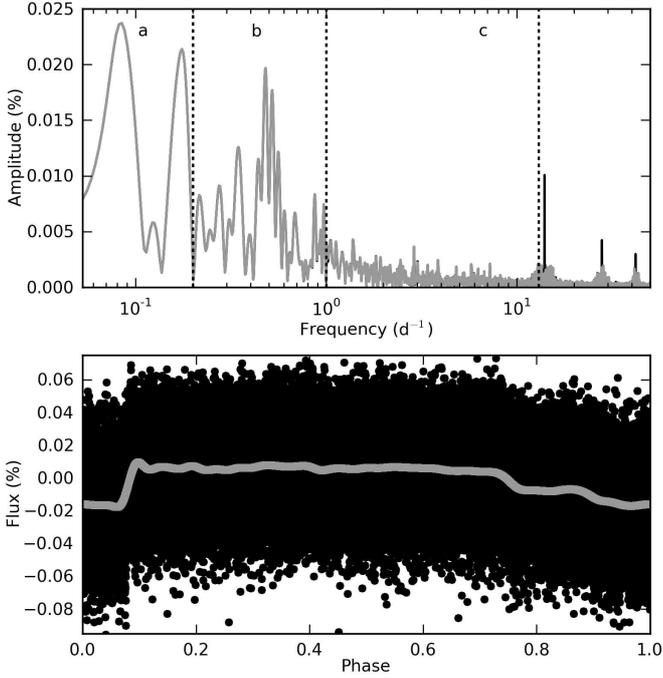}
\caption{(\emph{top panel}) Fourier periodogram of the CoRoT light curve of HD\,46149, with
the global trend removed (black) and with additional filtering of the satellite's
orbital influences (grey). The black line is only visible where the grey and black
lines do not overlap. The vertical dashed lines indicate the three different
regions (a, b, c) in which stellar signal is detected. (\emph{bottom panel}) The residual CoRoT light
curve where only the signal from the satellite is retained, folded on the
satellite's orbital frequency ($\approx 13.97$\,d$^{-1}$). The solid grey line
represents the spline model for the instrumental signal.}
 \label{fig:remove_orbit}
\end{figure}

The SAA crossing is a semiperiodic event and causes the strongest aliasing effect in the
Fourier window, at $\approx 13.97$\,d$^{-1}$. This is coupled to a semiperiodic dip in
the light curve at the same position, introducing an artefact that is related both
to sampling and flux variations. Because of the semiperiodic nature of the event,
it is difficult to completely remove it from the light curve. However, since the
nonharmonic signal with nearly constant frequency is
heavily confined to specific bandwidths at regular intervals, we minimised
the effect by first converting the light curve to Fourier space by iterative linear prewhitening,
and then removing all signals related to frequencies outside the interval $13.97\pm0.05$\,d$^{-1}$. We assumed
that the remainder of the signal is only due to the orbit of the satellite, and removed
that signal by fitting a 50th order spline of 3rd degree to the phasediagram of the
dominant frequency, determined with the phase dispersion minimization procedure \citep{stellingwerf1978} to
account for the highly non-sinusoidal form of the signal. The parameters for the
spline fit were fixed empirically, to optimally capture the discontinuity with
a relatively low number of knot points. The raw Fourier periodogram
with the orbital influence still present, the orbit model, and Fourier periodogram
of the cleaned signal are shown in Fig.\,\ref{fig:remove_orbit}.

\begin{figure}
\includegraphics[width=\columnwidth]{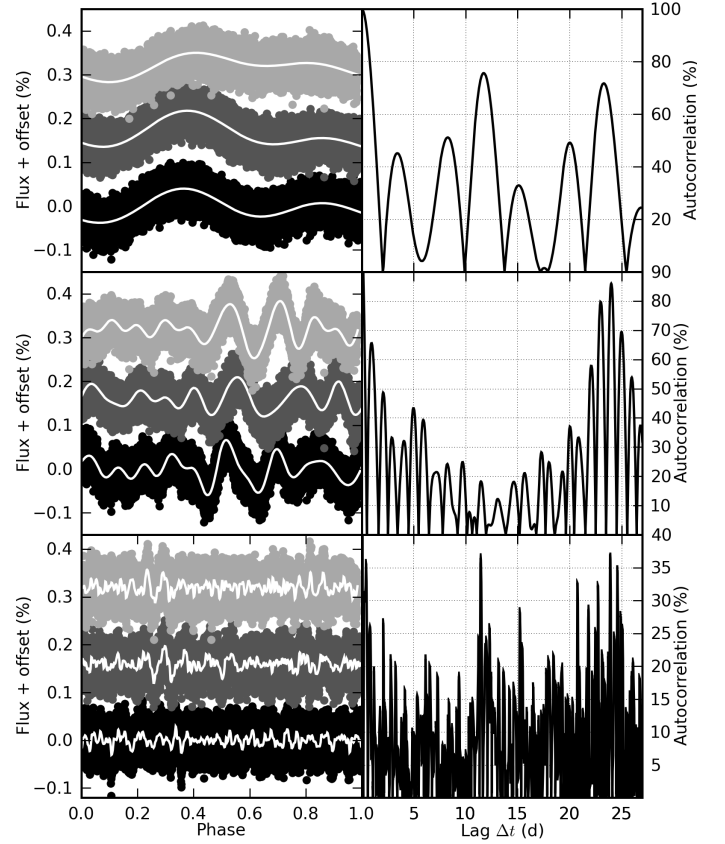}
\caption{(\emph{left panels}) CoRoT light curves folded on the period
$P=11.7$\,d, where all the signal outside three adjacent passbands is removed
(defined in Fig.\,\ref{fig:remove_orbit}). Consecutive phases are stacked on top
of each other with increasing offset, and shown in consecutive shades of grey. (\emph{top to bottom}) Low-frequency signal, mid-frequency
signal, high-frequency signal. (\emph{right panels}) Autocorrelation functions of
the filtered light curves in the corresponding left panel.}
 \label{fig:autocorrelation_phases}
\end{figure}

On the fully reduced light curve, we performed a traditional iterative prewhitening
procedure \citep[see, e.g.,][]{degroote2009b}. Because the first phase
diagram has a non-sinusoidal shape, we did not perform a nonlinear fitting procedure
using the inadequate sum-of-sines model. Instead, we used the Fourier decomposition to filter out specific bandwidths, between
0 and 0.2\,d$^{-1}$, 0.2 and 1.0\,d$^{-1}$, and 1.0 and 13\,d$^{-1}$ (regions a, b, c in Fig.\,\ref{fig:remove_orbit}).
A common feature of the isolated signal from these regions is their self similarity:
the autocorrelation of the low-frequency region reaches the $75\%$ level at
$\Delta t_l=11.76\pm0.96$\,d, the mid-frequency region shows a similar pattern
when folded on this period (although the autocorrelation attains a maximum at $2\Delta t_l$),
and a similar period is recovered in the highest peak in the autocorrelation of the
high-frequency signal ($\Delta t_h= 11.5\pm0.1$\,d) although at much lower amplitude ($36\%$).
The resemblance is confirmed by eye from the phase folded
light curves with $P=11.7$\,d (Fig.\,\ref{fig:autocorrelation_phases}).

\subsection{The frequency spacing}

\begin{figure}
\includegraphics[width=\columnwidth]{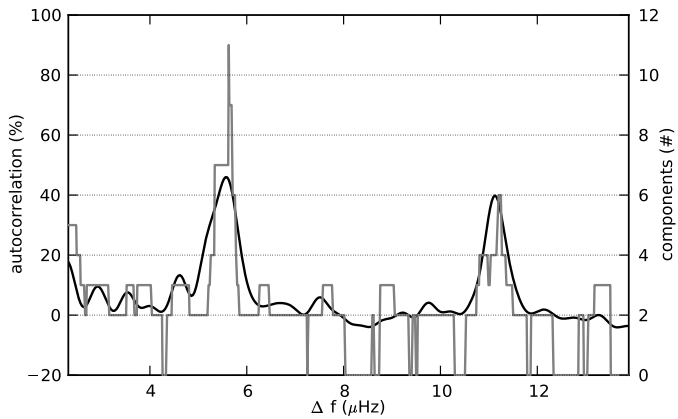}
\caption{Autocorrelation function of the periodogram of the light curve where
all low and mid-frequency signals are removed (black solid line), and the number
of components in a chain of corresponding frequency spacing $\Delta f$ (grey line).}
 \label{fig:freqspacing}
\end{figure}

After prewhitening the long period signal below 3\,d$^{-1}$, a prominent frequency spacing is detected in the high-frequency regime. The limit of 3\,d$^{-1}$ is chosen because
no dominant higher frequency peaks exist that can contaminate the lower
amplitude peaks. Both the autocorrelation of the periodogram
and spacing detection algorithms \citep{degroote2009a} uncover a spacing of
$\Delta f=0.48\pm0.02$\,d$^{-1}$, where at least 12 individual members of the spacing
can be recovered in the original linear frequency analysis (Fig.\,\ref{fig:freqspacing}).

\begin{figure}
\includegraphics[width=\columnwidth]{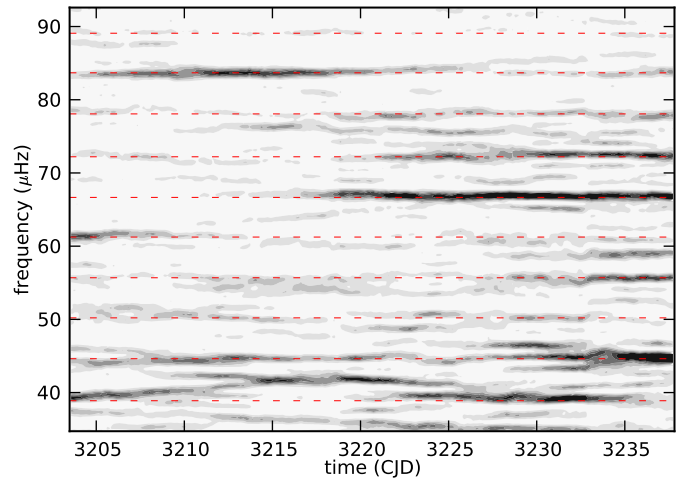}
\caption{Short time Fourier transformation (window width=10\,d) of the high-frequency region of the light curve, after filtering out the low- and mid-frequency regions. The different members of the frequency spacing identified from the complete light curve are indicated with red dashed lines, and show that the amplitudes connected to these frequencies vary significantly over the course of the time series ($t_0=2451545$\,HJD).}
 \label{fig:stochastic}
\end{figure}

\begin{figure}
\includegraphics[width=\columnwidth]{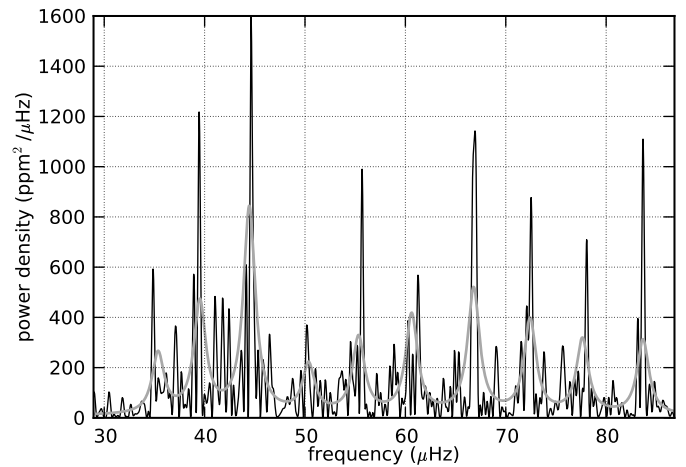}
\caption{Observed power spectrum (black line) and Lorentzian fit (grey line) with
the same lifetime for all frequency peaks. The width of the Lorentzian profile
implies a mode lifetime of the order of hours.}
 \label{fig:lorenz}
\end{figure}

Fourier transformations of selected pieces of the light curve are unable to reproduce
the values of the amplitudes. Specifically, a short-time Fourier transformation
is not able to trace any of the frequencies throughout the entire light curve, and the emerging pattern suggests a stochastic nature for the modes (Fig.\,\ref{fig:stochastic}). Since the
frequency resolution is high enough to adequately separate the different components,
this implies that the modes are excited and damped, bearing a resemblance to \emph{solar p-modes}  \citep{baudin1994}. Under such conditions, the mode lifetime can be estimated by fitting
Lorentzian profiles simultaneously to $N$ peaks in the power density spectrum $P(f)$ \citep[e.g.,][]{appourchaux1998,carrier2010},
via
\[P(f) = \sum_{n=1}^N\left(\frac{H_n}{1 + \left(\frac{2(f-f_n)^2}{\Gamma}\right)}\right) + B.\]
In this equation, $H_n$ is the height of the profile of frequency $f_n$, B is
the noise level, and $\Gamma$ the mode line-width at half maximum. The power spectral density is obtained by multiplying the power spectrum with the total time span $T$. The fit is then
performed by minimising
\[F = \ln P(f) + \frac{P(f)}{P_{fit}(f)},\]
using the Levenberg-Marquardt minimisation algorithm (Fig.\,\ref{fig:lorenz}). We decided to fix the mode lifetime to be equal for all the modes, reducing the number of parameters to fit, but removing the ability to constrain mode lifetimes individually. From the fit, we infer an average mode line width of $\Gamma = 1.05\pm0.15$\,$\mu$Hz corresponding to an average mode lifetime of $3.5^{+0.6}_{-0.4}$ days, which is consistent with the lifetime of the features in the short time Fourier transformations (Fig.\,\ref{fig:stochastic}). The fitted mode heights, amplitudes, and frequencies are listed in Table\,\ref{tbl:lorenz}, where the mode amplitude $A_{\rm rms}=\pi H\Gamma$. These amplitudes are not bolometric amplitudes but measured within the CoRoT bandpass \citep{michel2009}.

The error estimates in Table\,\ref{tbl:lorenz} are derived using Monte Carlo simulations. Thousands of different realisations of the light curve were constructed by adding white Gaussian noise to the light curve after prewhitening the long period signal, where the standard deviation $\sigma$ is determined as $\sigma=\sigma'/\sqrt{2}$, with $\sigma'$ the standard deviation of light curve after differentiating every two consecutive points. Adding the extra noise to the original light curve results in an increase in the true noise level, making the error estimates conservative.

Since no spectroscopic or multi-colour mode identification is possible for any
of the spaced frequencies, we cannot use the traditional forward modelling approach  for low-degree modes \citep[e.g.]{ausseloos2004} to compare theoretical models with the observations. Instead of fitting the frequency values separately, we use the method commonly adopted for solar-like oscillations, where we search for correspondence of both the value of the large frequency spacing, and the range of frequencies where the spacing occurs. Mode identification can then be done using the location of the ridges in the \'echelle diagram. In doing so, we want to establish a proof-of-concept, rather than full modelling of the stellar interior.

Before the observed spacing can be compared with stellar models, we have to constrain the star's fundamental parameters. Since we are dealing with a binary system for which no orbital constraints are given in the literature, we first concentrate on the binarity, to characterise the two components.

\begin{table}
 \caption{Parameters of Lorentzian fits to the frequency spectrum ($\Gamma=1.05$\,$\mu$Hz)}
\centering\begin{tabular}{cccr}
\hline
$f_n$\,($\mu$Hz) & $f_n$\,(d$^{-1}$) & $H_n$\,(ppm$^2$/$\mu$Hz) & $A_{\rm rms}$ (ppm)\\\hline
35.27 (0.15)       & 3.05              & 229 (38)                 & 27.0 (2.9)\\
39.54 (0.12)       & 3.42              & 435 (66)                 & 37.3 (4.0)\\
44.49 (0.11)       & 3.85              & 819 (102)                & 50.9 (3.9)\\
50.34 (0.15)       & 4.35              & 210 (61)                 & 25.3 (3.3)\\
55.37 (0.09)       & 4.79              & 332 (68)                 & 32.4 (3.0)\\
60.70 (0.16)       & 5.25              & 389 (66)                 & 35.2 (3.0)\\
66.81 (0.06)       & 5.77              & 512 (67)                 & 40.6 (2.9)\\
72.43 (0.08)       & 6.26              & 391 (67)                 & 35.2 (2.9)\\
77.70 (0.15)       & 6.71              & 293 (49)                 & 30.7 (2.9)\\
83.66 (0.04)       & 7.23              & 308 (36)                 & 31.5 (2.7)\\\hline
\end{tabular}
\label{tbl:lorenz}
\end{table}

\section{Orbital and fundamental parameters}\label{sec:orbital}
\subsection{Orbital parameters}
To construct the radial velocity curve, we used the measurements from \citet{mahy2009}. They were obtained by averaging the fitted
minimum of a Gaussian profile to the bottom halfs of selected He{\sc I} lines from
high-resolution spectra. To this set of observations, we added high-resolution spectra taken
with the \textsc{Coralie} \citep[$R\approx 50\,000$][]{coralie} and \textsc{Hermes} \citep[$R\approx 85\,000$)][]{hermes} spectrographs on the 1.2m twin telescopes Euler (La Silla, Chile)
and Mercator (La Palma, Spain). Part of the \textsc{Coralie} spectra were taken simultaneously with the CoRoT light curve. To account for the low signal-to-noise ratio
in these spectra (S/N of $\sim60$), the RV were obtained by two independent methods. First,
we fitted Gaussian profiles to the He{\sc I} $4471$\,\AA, He{\sc I} $5875$\,\AA, O{\sc III} $5592$\,\AA\ and
Si{\sc IV} $4088$\,\AA\ absorption lines and considered the
difference of the Gaussian minimum and the rest wavelength\footnote{Ralchenko, Yu., Kramida, A.E., Reader, J. and NIST ASD Team (2008). NIST Atomic Spectra Database. Available: http://physics.nist.gov/asd3. National Institute of Standards and Technology, Gaithersburg, MD.} as a measure for the
radial velocity. Because the upper part of some of these lines
show a significant departure from a Gaussian profile (because of blending or additional
broadening mechanisms), we only used the lower part
of the lines to fit the Gaussian profile. The exact cutoff depth $C$ was determined from
a trade-off between the reduced $\chi^2$ of the fit and the number of points used in the
fit via
\[ S(C) = 1/N^2 \sum \frac{(O_i-F_i)^2}{\sigma^2},\]
where $O_i$ and $F_i$ are the observed and fitted spectrum below the normalised
continuum level $C$, $\sigma^2$ is the variance, and a minimum in $S$ was pursued. There are consistent offsets
in the RV determination from the different lines of $\sim 5$~km\,s$^{-1}$.

Second, we computed the cross correlation function (CCF), using 8 absorption
lines with a minimum below 90\% of the continuum flux level between 4300 and $4800$\,\AA, while
removing the continuum to suppress the noise.
Since we have no appropriate template spectrum, we used the first observed
spectrum as a template, losing the ability to calibrate the RV in an absolute way.
Instead, we compared the differences in RV estimated with both methods, and
concluded that they were consistent with each other within $\sim$5 km s$^{-1}$. As a final
value for the RVs, we used the average of the estimations from the Gaussian line
profiles, and adopted the spread as a measure for the error.

Next, 2 high-resolution spectra were added from the \textsc{Harps} spectrograph on the La Silla 3.6m telescope \citep{harps}, from which the RV were derived in the same manner as described above.
Finally, we added the single radial velocity measurement from \citet{underhill1990},
which was determined via a line-bisector method using a single absorption line.

\begin{figure}
 \includegraphics[width=\columnwidth]{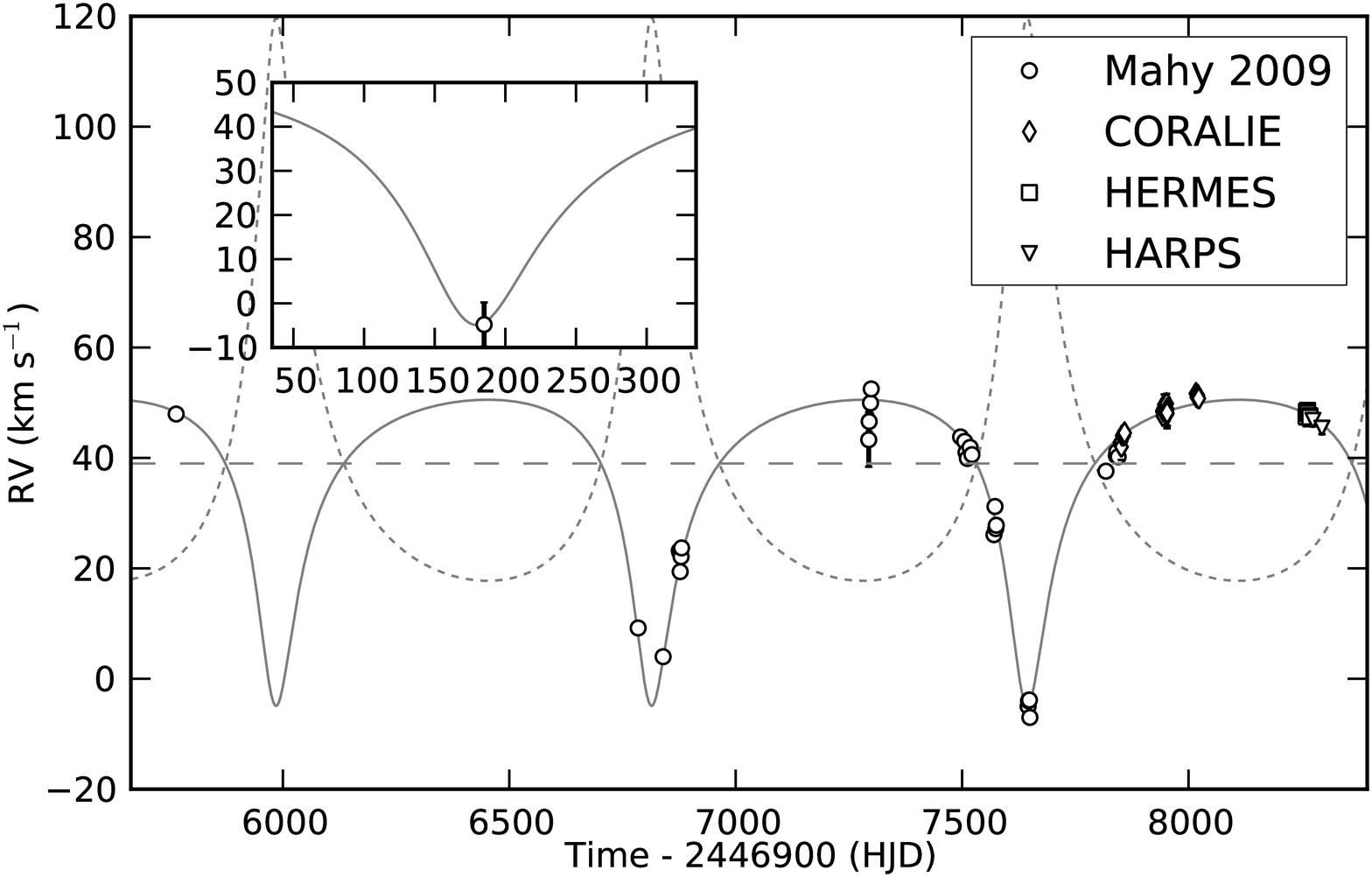}
\caption{Best Keplerian orbit fit (solid grey line) of the binary system HD\,46149
based on archival RV measurements from \citet{mahy2009}, and on new \textsc{Coralie}, \textsc{Hermes}, and \textsc{Harps} spectra. The short dashed line represents the predicted orbit of the
companion, corresponding to the spectroscopic fundamental parameters listed in Table\,\ref{tbl:fundpars}. The long dashed grey line represents the mean radial velocity $\gamma$ of the system. Inset is a zoom on the RV measurement of \citet{underhill1990}, which was done $\sim$15 years before the first measurement of \citet{mahy2009}.}\label{fig:orbit}
\end{figure}

Because the secondary component in HD\,46149 is only marginally visible (Fig.\,\ref{fig:fundpars}), we first treated it as a single-lined binary. Therefore, we fitted the best Keplerian orbit to the single set of RVs with the generalised least-square method of \citet{zechmeister2009}, and improved them
using a nonlinear fitting algorithm \citep{press1988} (Fig.\,\ref{fig:orbit}). The parameter estimations
of this long-period, highly eccentric system are listed in the top part of Table\,\ref{tbl:fundpars}.
The errors on the parameters were estimated using three methods; (a) bootstrapping 1000 random samples from the observations, (b) comparing the fitted parameters with the values obtained via a weighted fit, where we artificially reduced the influence
of the \textsc{Coralie} and \textsc{Hermes} measurements, since they are high in number but have a
narrow spread in time. The third method (c) was a Monte Carlo simulation,
generating thousands of datasets from the observed values and their estimated
uncertainties. Per parameter, the largest error estimate obtained by the three
methods was adopted, and are listed in Table\,\ref{tbl:fundpars}.

The mass function for single-lined spectroscopic binaries \citep{hilditch2001} follows from the eccentricity, semi-amplitude, and period listed Table\,\ref{tbl:fundpars}, and is equal to
\begin{equation}f(M_1,M_2,i) =\frac{(M_2\sin i)^3}{(M_1+M_2)^2} = 0.96\pm0.06.\label{eq:sb1}\end{equation}
Assuming an orbital plane perpendicular to the plane of the sky ($i=90^\circ$),
and a lower limit on the mass of primary of $M_1\geq20 M_\odot$, we arrive at a lower limit on the mass of the secondary of $M_2\geq 9.3M_\odot$.

\subsection{Fundamental parameters}\label{sec:fundpars}
\begin{figure}
 \includegraphics[width=\columnwidth]{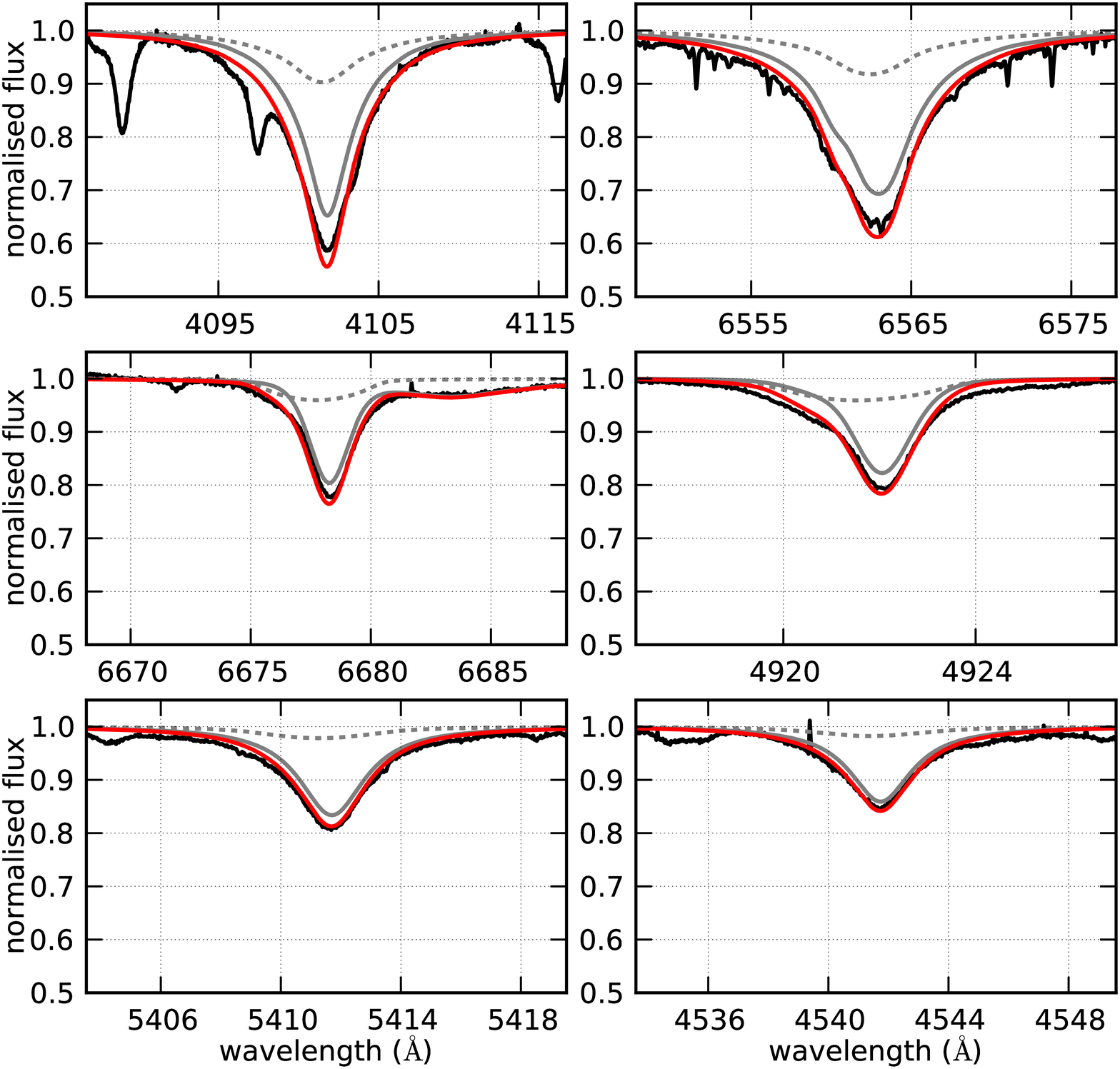}
\caption{Fits to the averaged \textsc{Coralie} spectra (black). The parameters
are listed in Table\,\ref{tbl:fundpars}. The fit to the primary companion is shown
in solid grey, the fit to the secondary in dashed grey, and the combined fit in
red. The H$\delta$ and H$\alpha$ lines are most sensitive to
the gravity, and $H\alpha$ does not show any signs of a wind \emph{(top panels)}.
In the combined He{\sc I}6678+He{\sc II}6683 and He{\sc I}4922 plots, the companion is clearly visible in the blue wing \emph{(middle panels)}. The He{\sc II}5410 and He{\sc II}4541 are the most sensitive to the difference in temperatures of both components \emph{(bottom panels)}.}
\label{fig:fundpars}
\end{figure}

The spectral type of HD\,46149 estimated by different authors ranges between O8V((f)) and O8.5V.
For better establishing the stellar parameters of the primary component (and obtaining a rough estimation of the
parameters of the secondary), we performed a spectroscopic analysis of the averaged \textsc{Coralie} spectrum
by means of the stellar atmosphere code FASTWIND \citep{puls2005}, because these turned
out to have the highest combined S/N. We applied the standard procedure in
which a set of H and He{\sc I}-{\sc II} lines in the optical spectra is fitted with synthetic lines from a grid of stellar
atmosphere models created to this aim. In the averaging, binarity was taken into account by summing
the profiles and assuming different luminosity ratios ($F_1/(F_1+F_2)\in [0.5,1]$). The projected equatorial rotation
velocity $v_{eq}\sin i=30\pm10$\,km\,s$^{-1}$ of the primary was obtained by applying the Fourier transform method \citep[][see also \citet{simondiaz2007} for a recent application to OB-type stars]{gray1992}
to the O{\sc III} 5592\,\AA\ line. The $v_{eq}\sin i$ of the secondary was considered as a free parameter in the analysis. The parameters of the best fit and corresponding model
parameters are listed in the top part of Table\,\ref{tbl:fundpars}, where the uncertainties should
not be strictly interpreted as the normal standard deviation, since many of the
underlying distributions are skewed or are unknown. The spectroscopic solutions
of both components lie on model isochrones, within the range expected from the age
of the cluster (Fig.\,\ref{fig:hrdiagram}). In Fig.\,\ref{fig:fundpars}, the overall
agreement is shown between the predicted and observed shape of the spectral lines. 

Once the spectroscopic parameters are fixed, we can use the profiles to estimate
the radial velocity of the companion, although the intrinsic uncertainty is high.
In a spectrum from \citet{mahy2009} taken at maximum velocity separation, we deduced
an RV of $\sim 110$\,km\,s$^{-1}$, which is compatible with the mass range deduced
from the effective temperature and gravity from the spectroscopic fit (Fig.\,\ref{fig:orbit}). Once the mass ranges of the two companions are fixed, we could deduce the inclination angle from Eq.\,(\ref{eq:sb1}), the semi-amplitude $K_2$ of the second component via $K_2 = (M_1 / M_2) K_1$, and the semi-major axes from the estimates of $a_{1,2}\sin i$ \citep[e.g.,][]{hilditch2001}. These are listed in the second part of Table\,\ref{tbl:fundpars}.

In the bottom part of Table\,\ref{tbl:fundpars}, we also list the radii, luminosities, and masses, fullfilling both the spectroscopic fundamental parameters and the parameters of the computed grid of stellar models. Finally, we note a discrepancy between the spectral energy distribution (SED) of the system and the derived fundamental parameters (see Appendix\,\ref{app:sed}).

\begin{table}
 \caption{Observed and corresponding model parameters.}
\centering\begin{tabular}{lcc}
\hline
Parameter & Value & Uncertainty \\
\hline
\multicolumn{3}{c}{from spectroscopy}\\\hline
$P$ (d)                        & $829$    & $4$ \\
$K_1$ (km\,s$^{-1}$)           & $27.7$   & $0.4$\\
$e$                            & $0.59$   & $0.02$\\
$\Omega$ ($\degr$)             & $172.1$  & $1.5$\\
$\gamma$ (km\,s$^{-1}$)        & $39.0$   & $0.3$\\
$T_0$ (HJD)                    & $2454538$& $5$\\
$T_{\rm eff,1}$ (K)            & $36000$  & $1000$ \\
$T_{\rm eff,2}$ (K)            & $33000$  & $1500$ \\
$\log_{10} g_1$ (cgs)          & $3.7$    & $0.1$ \\
$\log_{10} g_2$ (cgs)          & $4.0$    & $0.15$ \\
$v_{eq,1}\sin i$ (km\,s$^{-1}$)  & $30$     & $10$\\
$L_1/L_{\rm tot}$              & $0.75$   & --\\\hline
\multicolumn{3}{c}{from binarity (and spectroscopic values)}\\\hline
$i$\,$(\degr)$          & $49$     & $9$ \\
$a$ (AU)       & $6.5$    & $0.1$ \\
$K_2$ (km\,s$^{-1}$)           & $51$     & $9$ \\\hline
\multicolumn{3}{c}{from CL\'ES models (and spectroscopic values)}\\\hline
$R_1 (R_\odot)$                & $13$     & $2$\\
$R_2 (R_\odot)$                & $7.5$    & $2$\\
$\log_{10} L_1$ ($L_\odot$)    & $5.4$    & $0.2$\\
$\log_{10} L_2$ ($L_\odot$)    & $4.8$    & $0.2$\\
$L_1/L_{\rm tot}$              & $0.80$   & $0.20$\\
$M_1$ ($M_\odot$)              & $35$     & $6$ \\
$M_2$ ($M_\odot$)              & $19$     & $3$ \\
\hline
\end{tabular}
\label{tbl:fundpars}
\end{table}

\subsection{Spectroscopic variability}
Besides a small emission feature at CIII\,5696\,\AA\ (with an equivalent width of $22.5\pm0.5$\,m\AA),
there are no clear signs of emission in any of the spectra. We used
three different methods to search for line profile variability in a selection of
Si, O, He, and H lines: (a) we fitted a Gaussian profile to the lines and searched for variability in the moments \citep[e.g.,][]{aerts1992}, (b) we calculated
Fourier spectra per observed wavelength bin \citep[e.g.,][]{zima2006}, and (c) we
constructed bisectors and searched for variability at different normalised flux levels \citep[e.g.,][]{gray1992}. Except for the radial velocity shift due to the binary orbit, no significant
variability was detected.

\section{Discussion}\label{sec:discussion}

\subsection{Origin of the low-frequency variability}
The largest variations in the CoRoT light curve of HD\,46149 take place on long time scales of days. Because of the relatively short time span of the observations, this unavoidably means those features are only seen for a few cycles, so are poorly resolved. In Table\,\ref{app:freqlist}, we list the amplitudes and frequencies of the most prominent peaks, with a S/N above 4, calculated over a 6\,d$^{-1}$ interval in the periodogram before prewhitening. The listed error estimates are formal errors, assuming isolated peaks. Most of them, however, are relatively close to each other ($\Delta f\sim 1/T=0.03$\,d$^{-1}$), making it difficult to draw conclusions on the individual peaks.

One possible way of interpreting the low frequency peaks is that they originate from g mode pulsations. However, the first frequency $f_1=0.08373$\,d$^{-1}$ is clearly nonsinusoidal, because the first harmonic ($f_2=0.17506$\,d$^{-1}$) is also detected. From Fig.\,\ref{fig:autocorrelation_phases}, we see that these two frequencies generate a pattern of two consecutive `bumps' with different amplitudes, which is not a typical pulsation signature \citep[e.g.,][]{decat2002,decat2009}. For a nonlinear mode, a distorted phase shape is expected. Also from Fig.\,\ref{fig:autocorrelation_phases}, we see that the other high S/N peaks form a pattern that appears at fixed intervals in time, and vanish in between. The timescale between these patterns is similar to the period of $f_1$, $P=11.76$\,d. These considerations led us to conclude that the low-frequency signal is not likely to come from pulsations.

Another possible source of variability are atmospheric features, such as spots or chemically enhanced regions on the surface. The observed variability can then be explained by the appearance and disappearance of such features, either because of repeated creation and destruction or because of the rotation of the star. Here, we favour the second option, because of the signal's self-similarity, and because it is consistent with the spectroscopic determination of $v_{eq}\sin i=30\pm10$\,km\,s$^{-1}$. With this assumption and an estimate of the radius $R=13\pm2$\,$R_\odot$ (see Table\,\ref{tbl:fundpars}), we derive an inclination angle
\[i = \arcsin \left(\frac{v_{eq}\sin i}{2\pi R\Omega}\right) \approx 32\degr,\]
with a lower limit of $20\degr$. The ambiguity concerning the possibility that $2P$
is the true rotation period of the star is resolved by this argument, because
no inclination angle is able to explain such a low projected rotational velocity. We note the strong discrepancy up to a factor of two, with the projected rotational velocity of $\sim70$\,km\,s$^{-1}$ from, e.g., \citet{uesugi1970}. These authors assumed the whole broadening of the line profiles to be caused by rotation, while several studies have
shown that this hypothesis may be incorrect in O and B stars \citep[see, e.g.,][]{ryans2002,simondiaz2007}.
In contrast to previous estimations, we used the Fourier transform technique which,
in principle, allows separating rotational broadening from other non-rotational
broadenings in OB-type stars \citep{simondiaz2007}.

An important source of spectroscopic line profile variations in O stars is cyclical
wind variability \citep[e.g., ][]{henrichs2005}. Due to having only one weak
emission line (CIII\,5696), which is thought to be formed by photospheric
overpopulation \citep{leparskas1979},
and the absence of H$\alpha$ emission, we deduce that HD\,46149 has
a very weak wind, in agreement with the result of \citet{garmany1981}. Although
the S/N is too low to detect periodic variations in the spectra, there are faint
hints that the wind is not constant. This is seen in the higher noise level of
the amplitudes in the red wing of the hydrogen profiles than in the
blue wing.
In contrast, it is possible to detect the accumulated effect of all line profile
variations in the broad-band integrated photometry performed by the CoRoT satellite.
The predicted observed variation in the equivalent width of a specific spectral line
to account for, e.g., a 0.02\% amplitude in the light curve can be calculated as follows.
First, a normalised observed spectrum is rescaled to a \textsc{Fastwind} model atmosphere of $T_{\rm eff}=35\,000$\,K
and $\log g=4.0$. Next, the calibrated spectrum is multiplied by the response curve
of the CoRoT Astero-channel and normalised to have unit area. Finally,
the equivalent line width variation of the model is predicted by rescaling the desired
flux percentage to the normalised flux level. These calculations predict that an EW variation of 0.94\,\AA\ at 4000\,\AA, 0.67\,\AA\ at 5000\,\AA, and 1.14\,\AA\ at 6000\,\AA\ would separately all amount to a 0.02\% integrated flux change. The influence on one specific line is much less, since the EW variation
is spread over many lines. Given these numbers and the consideration
that the flux continuum is also susceptible to variability, it is no surprise
that we were unable to detect variability in the sparse timeseries of low S/N
spectra.

\subsection{The frequency spacing}\label{sect:freqspacing}

\begin{figure}
\includegraphics[width=\columnwidth]{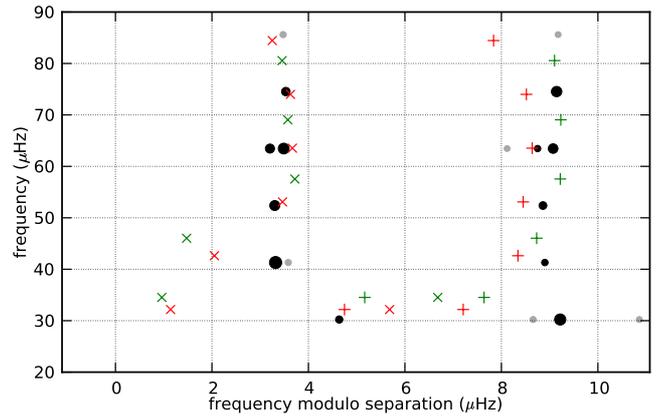}
\caption{\'Echelle diagram of observed frequencies (filled circles, the size of the
dot scales with the power). Black circles represent modes with an amplitude
above 11.5 ppm, the grey circles are modes with an amplitude above 10.5 ppm. The
frequency values of $\ell=0$ ($\times$) and $\ell=1$ ($+$) modes are overplotted
for the models of the primary component listed in Table\,\ref{tbl:models}
(red symbols denote the 30\,$M_\odot$ model,
green symbols denote the 34\,$M_\odot$ model).}
\label{fig:echelle}
\end{figure}

From the \'echelle diagram (Fig.\,\ref{fig:echelle}), we can see the similarity
with stochastically excited solar-like oscillations in low-mass stars,
which are interpreted as pressure modes in the asymptotic frequency regime. From an
initial extrapolation of the scaling law of \citet{kjeldsen1995}, the
observed separation of $\Delta f=0.48$\,d$^{-1}$ (or $5.5$\,$\mu$Hz) is too small to be the large separation, and too large to be the small separation. To investigate this discrepancy further, we used the grid of stellar evolution models.

Starting from the list of adiabatic frequencies for these models, we searched for frequency spacings in the $\ell=0,1$
modes between $\Delta f=0.2$\,d$^{-1}$ (2.3\,$\mu$Hz) and $\Delta f=3$\,d$^{-1}$ (34.7\,$\mu$Hz), where a small deviation of 0.1\,d$^{-1}$ was allowed from exact equidistance. We noticed that the extrapolated predicted frequency spacing from \citet{kjeldsen1995},
\[\Delta f = 134.9 \sqrt{\frac{M}{R^3}}\,\mu\mbox{Hz},\]
is only valid for models near the terminal-age main sequence (TAMS) where the spacing value is low. The model grid used in this paper, gives a slightly lower constant of proportionality of $120.4\pm1.4$ for massive stars above 20\,$M_\odot$. We note, however, that the calculated frequencies are low-order p-modes ($n\leq10$), so the law of \citet{kjeldsen1995} might not be applicable in this
case.

\begin{figure}
 \includegraphics[width=\columnwidth]{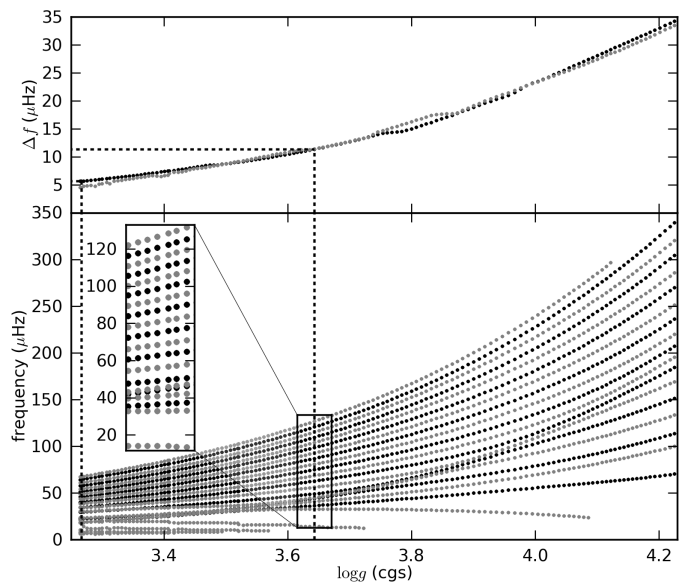}
\caption{\emph{Lower panel:} Evolution of the adiabatic eigenfrequencies of a 30\,$M_\odot$ model during
the main sequence ($\ell=0$ modes are plotted in black, $\ell=1$ in grey). \emph{Upper panel:} the value of the median frequency spacing. Dashed lines represent the frequency spacing $\Delta f=0.48$\,d$^{-1}$ and $2\Delta f$. Inset is a zoom on the region compatible with the spectroscopic determination of the gravity.}
\label{fig:evolution}
\end{figure}

From the spacing search analysis of the CL\'ES models, we deduced that, for
low-degree modes, $\Delta f$ is above 3\,d$^{-1}$ at the zero-age main sequence,
and gradually decreases to values around 1\,d$^{-1}$ and lower as the star evolves towards the TAMS, because of the increase in radius (Fig.\,\ref{fig:evolution}). At the same time, the lower limit of the eigenfrequencies where the spacings occur, decreases from $\sim$25\,d$^{-1}$ (290\,$\mu$Hz) to $\sim$5\,d$^{-1}$ (58\,$\mu$Hz).

\begin{figure}
\includegraphics[width=\columnwidth]{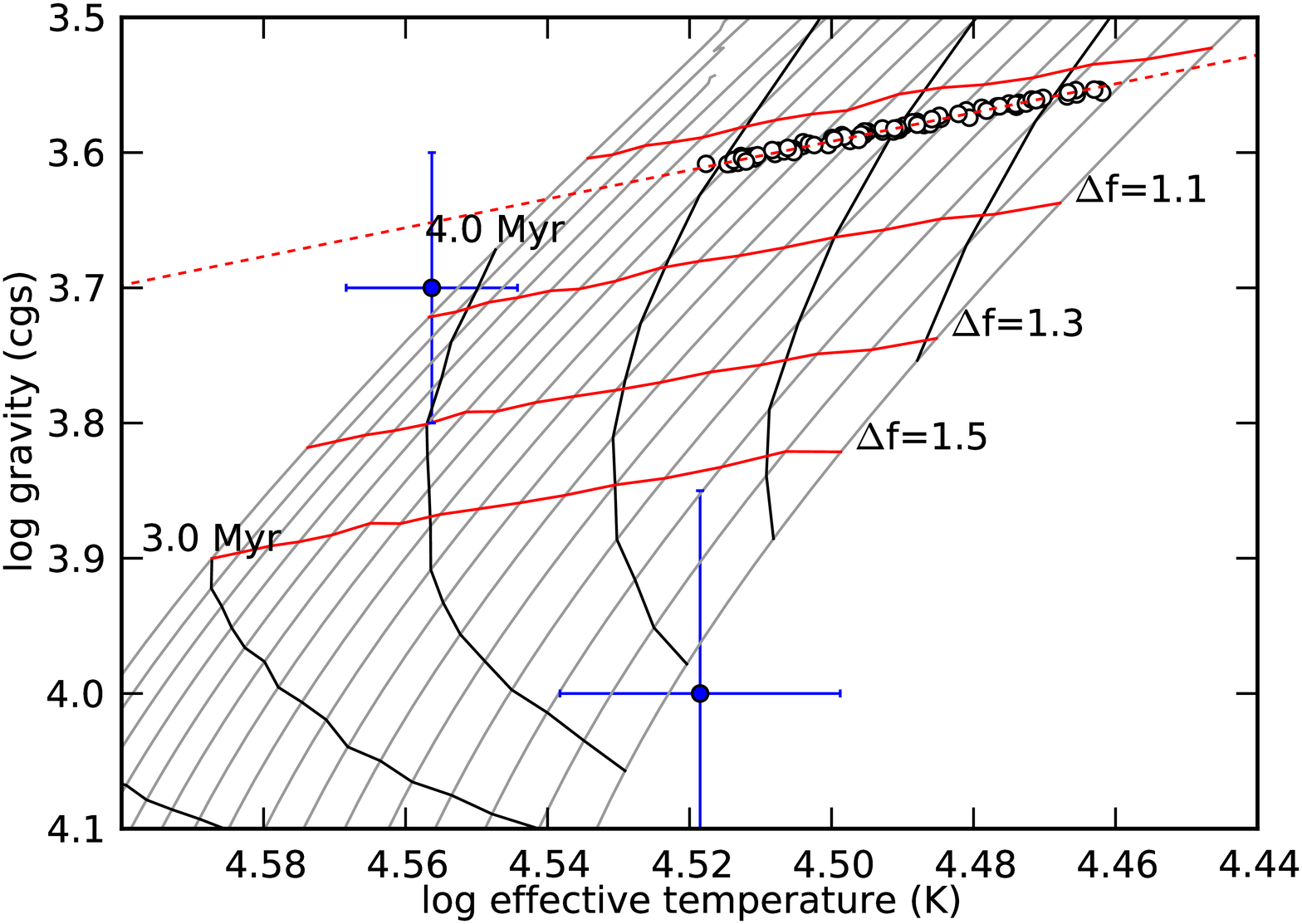}
\caption{$\log\,T_{\rm eff}$-$\log g$ diagram of late O-type and
early B-type stars. The spectroscopic parameter and corresponding error estimations of both components
are indicated by the crosses. The evolutionary tracks (grey solid lines) are for masses
of 20-35\,$M_\odot$ and overshoot parameter $\alpha=0.2$. The black solid lines
denote the isochrones of 3-7\,Myr. The horizontal solid red lines show the lines of
constant density, and a large separation of $\Delta f=1.0,1.1,1.3$ and $1.5$. The open circles
indicate the position of the models best fitting the observed spacing ($\Delta f=0.98$). The
red dashed line is a linear fit through these positions. Some models above 30\,$M_\odot$
deviate from this fit, because the model grid is less dense in these regions.}
\label{fig:hrdiagram}
\end{figure}

We searched for the models in the grid for which both the spacing and the frequency values best match the observations. The best-fitting models were determined via a $\chi^2$-like statistic,
\[S = \frac{1}{N}\sum_i (f_{i,m}-f_{i,o})^2,\]
where a minimum in $S$ is searched for, resulting in those models where the average difference between the model frequencies and observed frequencies is minimal. From the location of the best-fitting models in the $\log\,T_{\rm eff}$-$\log g$ diagram
(Fig.\,\ref{fig:hrdiagram}), we inferred that they are located on a line of almost constant density, in agreement with what is known from the theory of stellar oscillations. We assumed that the observed spacing is half of
the large separation between $\ell=0$ modes (and thus the separation between $\ell=0$ and $\ell=1$) which is observed: a large separation of $\Delta f$ would mean the star is at the end of the main sequence (Fig.\,\ref{fig:evolution}), which is incompatible with spectroscopy. Thus, we assumed $\Delta f_{0,1}=0.48$\,d$^{-1}$. From this analysis, we inferred the following
relation between the effective temperature and gravity for models with the
observed spacing:
\[\log g = -1.20 + 1.06\log T_{\rm eff}.\]
Using the observed $T_{\rm eff}=36000\pm1000$\,K, the latter fit predicts a
gravity of $\log g=3.63\pm0.02$ (cgs), which is compatible with the observed value
of $\log g=3.7\pm0.1$ (cgs) of the primary component of HD\,46149. We compared the
calculated frequency spectrum of two models for the primary on the $\log g-T_{\rm eff}$
line defined above, with the observed frequency spectrum in Fig.\,\ref{fig:echelle}. The
parameters of the models are listed in Table\,\ref{tbl:models}. Overall, there is good agreement between the location of the ridges at high frequency, although the frequency values do not fit well individually. At lower frequencies, the observed frequencies deviate significantly from the model frequencies. We attribute this to the fact that, in the models, the spacing only appears above $\sim50$\,$\mu$Hz (Fig.\,\ref{fig:evolution}), making the low frequencies more sensitive to other parameters than the global mean density. In a future, more in-depth study of the pulsational characteristics of HD\,46149, the discrepancies between the observed and theoretical frequency values should be accounted for. The lowest frequencies in the spacing could add additional constraints on the model physics.

\begin{table}
\caption{Two representative models with overshoot parameter $\alpha=0.2$ for the
primary component of HD\,46149, which show the observed spacing between the
$\ell=0,1$ modes.}
\begin{tabular}{cccccc}
\hline
$M/M_\odot$ & $\log T_{\rm eff}$ &$\log g$ & $R/R_\odot$ & $\log L/L_\odot$ & Age (Myr) \\\hline
30          & 4.512              & 3.575   & 14.80       & 5.341            & 5.0\\
34          & 4.539              & 3.643   & 14.56       & 5.436            & 4.2\\
 \hline

\end{tabular}
\label{tbl:models}

\end{table}

\section{Conclusions}
We discovered modes with a finite lifetime in a massive O star binary system, for
which we determined the orbital parameters, and constrained the fundamental
parameters of both components spectroscopically. We removed the large-scale
photometric variations, which we argued are due to changing features in the stellar
atmosphere compatible with the rotational cycle of the star, instead of pulsations. The exact origin of this
variability is unknown, but could be attributed to spots, stellar winds, or
chemical inhomogeneities. After removing of these features, we interpreted the remaining signal as low-order pressure modes
with a finite lifetime. Similar modes
have been claimed before in the massive $\beta$\,Cephei star HD\,180642 by
\citet{belkacem2009}, and were interpreted as solar-like oscillations, excited by
the convective region induced by the iron opacity bump \citep{belkacem2010}.
In the case of HD\,46149, the frequency separation is compatible with the
characteristic spacing between $\ell=0,1$ modes in stellar models. The observed spacing led to a mean density of the star, which is in good agreement with the parameters of the system derived from spectroscopy.

In a follow-up study, in-depth seismic modelling of this very massive binary will be considered by computing a fine but extensive grid of stellar models with various choices of the input physics, to try and fine-tune the physics for the most massive stars.
Our observational results constitute a suitable
starting point for in-depth seismic modelling of this very massive binary.

\begin{acknowledgements}
We are grateful to Andrea Miglio, Josefina Montalban, M\'elanie Godart, and
Marc-Antoine Dupret of Li\`ege University for valuable discussions of the
theory of stellar oscillations of O stars. The research leading to these results has received funding from the European Research Council under the European Community's Seventh Framework Programme (FP7/2007--2013)/ERC grant agreement n$^\circ$227224 (PROSPERITY), as well as from the Research Council of K.U.Leuven grant agreement GOA/2008/04 and from the Belgian PRODEX Office under contract C90309: CoRoT Data Exploitation. SSD acknowledges financial support by the Spanish Ministerio de Ciencia e Innovaci\'on under the project AYA2008-06166-C03-01, and the Consolider-Ingenio 2010 Programme grant CSD2006-00070: First Science with the GTC  (http://www.iac.es/consolider-ingenio-gtc). The HARPS data are being obtained as part of the ESO Large Programme LP182.D-0356 (PI: E.\,Poretti). EP and MR acknowledge financial support from the Italian ESS Project, contract ASI/INAF I/015/07/0, WP03170. LM acknowledges financial support from the PRODEX XMM/Integral contract (Belspo).
\end{acknowledgements}

\bibliographystyle{aa}
\bibliography{14543}

\begin{appendix}
\section{Tables}

Table\,\ref{tbl:speclog} contains a list of the spectroscopic observations. The first column shows the time of observation, the second column the instrument used, and the third and fourth columns the exposure time and S/N, respectively.

Table\,\ref{tbl:sed} lists the literature photometry. The first column denotes the wavelength range or average wavelength of the response function. The second and third columns denote the photometric system and the name of filter. The last columns contain references to the catalogues used or instruments.

Table\,\ref{app:freqlist} is a list of all high S/N frequency peaks with their respective amplitudes. The S/N is computed by dividing the amplitude through the average noise level in the periodogram over a 6\,d$^{-1}$ interval in the periodogram. 

 \begin{table}
\caption{Logbook of spectroscopic observations.}
\centering\begin{tabular}{ccrr}\hline
time (HJD-2450000)    & instrument   & exp. time (s) & $\langle$ S/N$\rangle$\\\hline
4748.77151 & {\sc Coralie} & 1800 & 45\\
4750.80021 & {\sc Coralie} & 1800 & 56\\
4751.79249 & {\sc Coralie} & 1800 & 25\\
4753.83953 & {\sc Coralie} & 1800 & 62\\
4755.86699 & {\sc Coralie} & 1800 & 69\\
4756.75304 & {\sc Coralie} & 1800 & 57\\
4758.76337 & {\sc Coralie} & 1500 & 61\\
4842.59491 & {\sc Coralie} & 1800 & 64\\
4842.67488 & {\sc Coralie} & 1800 & 64\\
4842.75034 & {\sc Coralie} & 1800 & 61\\
4843.58534 & {\sc Coralie} & 1800 & 65\\
4843.64708 & {\sc Coralie} & 1800 & 62\\
4843.79525 & {\sc Coralie} & 1800 & 60\\
4844.62333 & {\sc Coralie} & 1800 & 63\\
4844.70675 & {\sc Coralie} & 1800 & 64\\
4845.59027 & {\sc Coralie} & 1800 & 56\\
4845.77435 & {\sc Coralie} & 1800 & 51\\
4846.58948 & {\sc Coralie} & 1801 & 64\\
4846.68792 & {\sc Coralie} & 1801 & 66\\
4846.74996 & {\sc Coralie} & 1801 & 64\\
4847.58828 & {\sc Coralie} & 1801 & 64\\
4847.68662 & {\sc Coralie} & 1801 & 60\\
4848.61953 & {\sc Coralie} & 1800 & 60\\
4848.69951 & {\sc Coralie} & 1800 & 62\\
4850.61492 & {\sc Coralie} & 1800 & 59\\
4850.67672 & {\sc Coralie} & 1800 & 59\\
4851.59201 & {\sc Coralie} & 1800 & 57\\
4852.56425 & {\sc Coralie} & 1800 & 49\\
4852.62809 & {\sc Coralie} & 1800 & 60\\
4852.73125 & {\sc Coralie} & 1800 & 58\\
4912.55844 & {\sc Coralie} & 1800 & 59\\
4916.51655 & {\sc Coralie} & 1800 & 60\\
4918.54146 & {\sc Coralie} & 1801 & 58\\
4920.55779 & {\sc Coralie} & 1801 & 59\\
4922.55891 & {\sc Coralie} & 1801 & 57\\
5159.72850 & {\sc Hermes} &  420 & 27\\
5160.47824 & {\sc Hermes} & 1200 & 65\\
5162.53878 & {\sc Hermes} &  800 & 63\\
5162.54759 & {\sc Hermes} &  600 & 57\\
5162.77441 & {\sc Hermes} &  800 & 64\\
5162.78425 & {\sc Hermes} &  800 & 60\\
5167.68048 & {\sc Hermes} &  900 & 45\\
5167.69148 & {\sc Hermes} &  900 & 44\\ 
5167.78661 & {\sc Hermes} & 1200 & 54\\
5174.68443 & {\sc Harps}  &  620 & 141\\
5194.66678 & {\sc Harps}  &  800 & 169\\
\hline\end{tabular}
\label{tbl:speclog}
\end{table}

\begin{table}
\caption{Literature photometry of HD\,46149.}
\begin{tabular}{rllr}
\hline
wavelength (\AA) & system & band & reference\\\hline
  900-1150 &   FUSE &     Far-UV &  \citet{fuse}\\
 1150-3300 &    IUE &         UV &  \citet{iue}\\
 2740 &        TD1 &         UV &  \citet{TD1}\\
 3641 &    JOHNSON &          U &  \citet{johnson}\\
 4448 &    JOHNSON &          B &  \citet{johnson}\\
 5504 &    JOHNSON &          V &  \citet{johnson}\\
12487 &    JOHNSON &          J &  \citet{johnson}\\
16464 &    JOHNSON &          H &  \citet{johnson}\\
21951 &    JOHNSON &          K &  \citet{johnson}\\
 4204 &      TYCHO &         BT &  \citet{tycho}\\
 5321 &      TYCHO &         VT &  \citet{tycho}\\
 4718 &       SDSS &          g &  \citet{sdss}\\
 6185 &       SDSS &          r &  \citet{sdss}\\
 7499 &       SDSS &          i &  \citet{sdss}\\
 8961 &       SDSS &          z &  \citet{sdss}\\
 7885 &    COUSINS &          I &  \citet{cousins}\\
12412 &      2MASS &          J &  \citet{2mass}\\
16497 &      2MASS &          H &  \citet{2mass}\\
21909 &      2MASS &         KS &  \citet{2mass}\\
35375 &       IRAC &         36$\mu$m &  \citet{irac}\\
44750 &       IRAC &         45$\mu$m &  \citet{irac}\\
57019 &       IRAC &         58$\mu$m &  \citet{irac}\\
77843 &       IRAC &         80$\mu$m &  \citet{irac}\\
87973 &        MSX &          A &  \citet{msx}\\\hline
\end{tabular}
\label{tbl:sed}
\end{table}

\begin{table}
\caption{List of high S/N frequency peaks}
\label{app:freqlist}
\centering\begin{tabular}{ccccr}\hline
 $A$\,(ppm) & $\sigma(A)$ & $f$\,($\mu$Hz) &  $\sigma(f)$ & S/N\\\hline
240.6 & 1.9 &  0.9691 & 0.0015 &  12.27  \\
199.6 & 1.8 &  2.0262 & 0.0015 &  11.91  \\
201.8 & 1.6 &  5.5546 & 0.0016 &  11.12  \\
163.1 & 1.5 &  6.0215 & 0.0017 &  10.33  \\
131.2 & 1.5 &  4.0575 & 0.0020 &   9.12  \\
 96.0 & 1.4 &  0.5159 & 0.0030 &   6.22  \\
 81.3 & 1.4 &  10.075 & 0.0032 &   6.01  \\
 85.0 & 1.4 &  6.4911 & 0.0030 &   6.34  \\
 88.6 & 1.3 &  5.1096 & 0.0031 &   6.37  \\
 83.5 & 1.3 &  7.9442 & 0.0030 &   6.61  \\
 63.9 & 1.3 & 11.2296 & 0.0035 &   5.87  \\
 64.3 & 1.3 &  1.7303 & 0.0037 &   5.75  \\
 57.0 & 1.2 &  3.6946 & 0.0044 &   4.93  \\
 52.8 & 1.2 &  2.9027 & 0.0046 &   4.71  \\
 47.0 & 1.2 &  3.2964 & 0.0050 &   4.50  \\
 51.7 & 1.2 & 12.1920 & 0.0050 &   4.57  \\
 41.7 & 1.2 &  7.0373 & 0.0053 &   4.38  \\
 49.0 & 1.2 & 15.9927 & 0.0056 &   4.24  \\
 33.6 & 1.2 & 10.8596 & 0.0059 &   4.04  \\
 36.6 & 1.2 & 33.4241 & 0.0060 &   4.01  \\
 19.0 & 1.1 & 83.6793 & 0.0106 &   4.17  \\\hline
\end{tabular}
\end{table}

\section{Spectral energy distribution}\label{app:sed}
\begin{figure}
\includegraphics[width=\columnwidth]{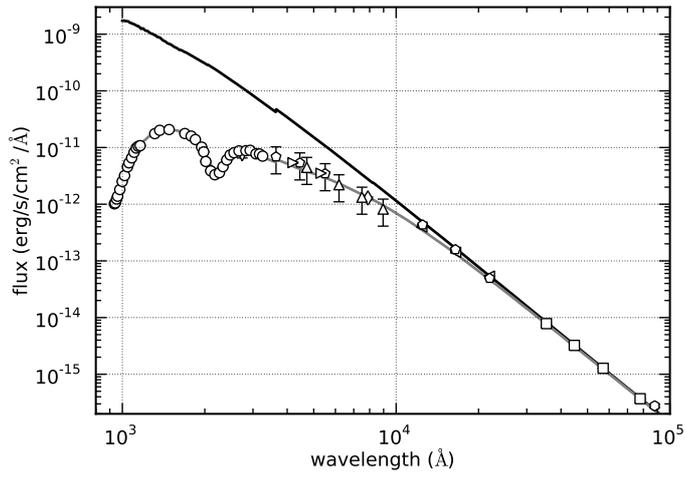}
\caption{Observed spectral energy distribution of the binary HD\,46149 (white symbols, for a summary,
see Table\,\ref{tbl:sed}) reddened
\textsc{Fastwind} model (grey solid line) with $E(B-V)=0.46$. The original model is shown
in black.}
 \label{fig:sed}
\end{figure}

The SED of the system can be used as an independent test for the derived fundamental parameters in Sect.\,\ref{sec:fundpars}. Some uncertainty is introduced, however, because the Balmer discontinuity is located in the
far UV part of the spectrum, which is potentially heavily distorted due to extinction
effects. Still, many flux measurements are found in the literature in a broad spectral
range (Table\,\ref{tbl:sed}), which can be used to infer that there is little or no 
dust surrounding HD\,46149, since no significant infrared excess is observed
(Fig.\,\ref{fig:sed}). The SED also delivers an independent measure for the radius of
the primary component, assuming a radius ratio of both components, via
\[F_{\rm obs}(\lambda) = \left(F_{\rm mod,1}(\lambda) +
F_{\rm mod,2}(\lambda)\frac{R_2^2}{R_1^2}\right)\frac{R_1^2}{d^2}E(\lambda).\]
A \textsc{Fastwind} \citep{puls2005} grid of model atmospheres was used to provide the
basic stellar flux models. The predicted fluxes
$F_{\rm mod,1,2}(\lambda)$ were computed by interpolating the grid to the values
obtained from spectroscopy (Table\,\ref{tbl:fundpars}), and the factor $R_2^2/R_1^2$
was also adopted from Table\,\ref{tbl:fundpars}. A grid search was performed to
find the minimum values of wavelength-dependent extinction $E(\lambda)$ \citep{cardelli1989} and $R_1^2/d^2$ in the following way: all values between $E(B-V)=0$ and $E(B-V)=2$ with a stepsize of 0.1 are adopted, and the search is iterated in a smaller interval around the minimum value until convergence is reached (i.e., until two consecutive minima differ less than 0.001 from each other). From the known distance to the system of $d=1.6\pm0.2$\,kpc
and fitted extinction $E(B-V)=0.46\pm0.05$ in agreement
with literature values, we derive the following relation between $R_1$
and $d$:
\[R_1 [R_\odot] = 5.13\ d\ \mbox{[kpc]}.\]
Thus we arrive at an estimate of the radius of $R_1=8.3\pm2$\,$R_\odot$, assuming
the radius ratio of 0.6 from Table\,\ref{tbl:fundpars}. This is lower than the one obtained from confronting models with the spectroscopic fundamental parameters. On the other hand, fixing the radius of the primary to the value in Table\,\ref{tbl:fundpars}, leads to overestimating the distance compared to the literature value. Such
discrepancies between different methods of parameter estimations have been reported
before \citep[e.g.,][]{herrero1992}.

\end{appendix}

\end{document}